\documentclass[MSNbibl,nameyear,dvips]{arxstspdf}
\usepackage{url,breakurl}
\usepackage{graphicx}
\usepackage{flushend}
\usepackage{stfloats}

\volume{29}
\issue{4}
\pubyear{2014}
\firstpage{662}
\lastpage{678}
\doi{10.1214/13-STS429} 

\makeatletter

\newcommand{\citepp}[1]{(\citeauthor{#1}, \citeyear{#1})}

\newcommand{\rright}{\right}
\newcommand{\lleft}{\left}
\newcommand{\rrvert}{\vert}
\newcommand{\llvert}{\vert}
\newtheorem{theorem}{Theorem}
\newtheorem{lemma}{Lemma}
\makeatother
\begin{document}
\begin{frontmatter}

\title{A Uniformly Consistent Estimator of Causal Effects under the
$k$-Triangle-Faithfulness Assumption}
\runtitle{A Uniformly Consistent Estimator of Causal Effects}

\begin{aug}
\author[a]{\fnms{Peter} \snm{Spirtes}\corref{}\ead[label=e1]{ps7z@andrew.cmu.edu}}
\and
\author[b]{\fnms{Jiji} \snm{Zhang}\ead[label=e2]{jijizhang@ln.edu.hk}}
\runauthor{P. Spirtes and J. Zhang}

\affiliation{Carnegie Mellon University and Lingnan University}

\address[a]{Peter Spirtes is Professor, Department of Philosophy,
Carnegie Mellon University,
5000 Forbes Avenue, Pittsburgh, Pennsylvania 15213, USA \printead{e1}.}
\address[b]{Jiji Zhang is Associate Professor, Department of
Philosophy, Lingnan University, Tuen Mun, N.T., Hong Kong \printead{e2}.}

\end{aug}

%
\begin{abstract}
Spirtes, Glymour and Scheines [\textit{Causation}, \textit{Prediction}, \textit{and Search} (1993) Springer]
described a pointwise consistent estimator of the
Markov equivalence class of any causal structure that can be
represented by a directed acyclic graph for any parametric family with
a uniformly consistent test of conditional independence, under the
Causal Markov and Causal Faithfulness assumptions.
Robins et~al. [\textit{Biometrika} \textbf{90} (2003) 491--515], however, proved that there are no uniformly consistent
estimators of Markov equivalence classes of causal structures under
those assumptions. Subsequently,
Kalisch and B{\"u}hlmann
[\textit{J. Mach. Learn. Res.} \textbf{8} (2007) 613--636]
described a uniformly consistent estimator of the Markov equivalence
class of a linear Gaussian causal structure under the Causal Markov and
Strong Causal Faithfulness assumptions. However, the Strong
Faithfulness assumption may be false with high probability in many
domains. We describe a uniformly consistent estimator of both the
Markov equivalence class of a linear Gaussian causal structure and the
identifiable structural coefficients in the Markov equivalence class
under the Causal Markov assumption and the considerably weaker
\textit{k}-Triangle-Faithfulness assumption.
\end{abstract}

%

\end{frontmatter}

\section{Introduction}\label{sec1}
A principal aim of many sciences is to model causal systems well enough
to provide sound insight into their structures and mechanisms and to
provide reliable predictions about the effects of policy interventions.
The modeling process is typically divided into two distinct phases: a
model specification phase in which some model (with free parameters) is
specified, and a parameter estimation and statistical testing phase in
which the free parameters of the specified model are estimated and
various hypotheses are put to a statistical test. Both model
specification and parameter estimation can fruitfully be thought of as
search problems.

As pointed out in \citet{Robetal03}, common statistical wisdom
dictates that causal effects cannot be consistently estimated from
observational studies alone unless one observes and adjusts for all
possible confounding variables, and knows the time order in which
events occurred. However, \citet{SpiGlySch93} and \citet{Pea00}
developed a framework in which causal relationships are represented by
edges in a directed acyclic graph. They also described asymptotically
consistent procedures for determining features of causal structure from
data even if we allow for the possibility of unobserved confounding
variables and/or an unknown time order, under two assumptions: the
Causal Markov assumption (roughly, given no unmeasured common causes,
each variable is independent of its noneffects conditional on its
direct causes) and the Causal Faithfulness assumption (all conditional
independence relations that hold in the distribution are entailed by
the Causal Markov assumption). Under these assumptions, the procedures
they propose (e.g., the \textit{SGS} and the \textit{PC} algorithms assuming
no unmeasured common causes, and the \textit{FCI} algorithm which does not
assume no unmeasured common causes) can infer the existence or absence
of causal relationships. In particular, Spirtes et al. (\citeyear{SpiGlySch93}), Chapters~5 and 6, proved the Fisher consistency of these procedures. Pointwise
consistency follows from the Fisher consistency and the uniform
consistency of the test procedures for conditional independence
relationships in certain parametric families that the procedures use.

\citet{Robetal03} proved that under the Causal Markov and
Faithfulness assumptions made in \citet{SpiGlySch93} there are no
uniformly consistent procedures for estimating features of the causal
structure from data, even when there are no unmeasured common causes.
\citet{SpiGlySch00}, \citet{KalBuh07} and \citet{Coletal12} introduced a Strong Causal Faithfulness assumption, which,
roughly speaking, assumes that no conditional independence relation not
entailed by the Causal Markov assumption ``almost'' holds. \citet{KalBuh07} and \citet{Coletal12} showed that under this
strengthened Causal Faithfulness assumption, some modifications of the
pointwise consistent procedures developed in \citet{SpiGlySch93} are
uniformly consistent. \citet{Maaetal10} have also successfully
applied these procedures to various biological data sets,
experimentally confirming some of the causal inferences made by the procedures.

However, the question remains whether the Strong Causal Faithfulness
assumption made by \citet{KalBuh07} is too strong. Is it
likely to be true? Some analysis done by \citet{Uhl12} indicates
that the strengthened Causal Faithfulness assumption is likely to be
false, especially when there are a large number of variables.

In this paper we investigate a number of different ways in which the
strengthened Causal Faithfulness assumption can be weakened, while
still retaining the guarantees of uniformly consistent estimation by
modifying the causal estimation procedures. It is not clear whether the
ways we propose to weaken the Strong Causal Faithfulness assumption
make it substantially more likely to hold, nor is it clear that all of
the modifications that we propose to the estimation procedures make
them substantially more accurate in practice. Nevertheless, we believe
that the modifications that we propose are a useful first step toward
investigating fruitful modifications of the Causal Faithfulness
assumption and causal estimation procedures.

In Section~\ref{sec2} we describe the basic setup and assumptions for causal
inference. In Section~\ref{sec3} we examine various ways to weaken the Causal
Faithfulness assumption and modifications of the estimation procedures
that preserve pointwise consistency. In Section~\ref{sec4} we examine weakening
the Strong Causal Faithfulness assumption and modification of the
estimation procedures that preserves uniform consistency. Finally, in
Section~\ref{sec5} we summarize the results and describe areas of future research.

\section{The Basic Assumptions for Causal Inference}\label{sec2}
We first introduce the graph terminology that we will use. Individual
variables are denoted with italicized capital letters, and sets of
variables are denoted with bold-faced capital letters. A \textit{graph} $
G = \langle\mathbf{V}, \mathbf{E} \rangle$ consists of a set of \textit{vertices} $\mathbf{V}$ and a set
of \textit{edges} $\mathbf{E} \subseteq\mathbf{V} \times\mathbf{V}$, where for each
$\langle X, Y\rangle\in\mathbf{E} ,\it{X} \neq\it{Y}$. If $
\langle X,Y\rangle\in\mathbf{E}$
and $\langle Y,X\rangle\in\mathbf{E}$, there is an \textit{undirected
edge} between \textit{X}
and \textit{Y}, denoted by \textit{X} --- \textit{Y}. If $\langle X,Y\rangle
\in\mathbf{E}$
and $\langle Y,X\rangle\notin\mathbf{E}$, there is a {directed edge}
between \textit{X}
and \textit{Y}, denoted by $X \rightarrow Y$. If there is a directed
edge from \textit{X} to \textit{Y}, or from
\textit{Y} to \textit{X}, or there is an undirected edge between \textit{X} and
\textit{Y}, then \textit{X} and \textit{Y} are \textit{adjacent} in \textit{G}. $
\operatorname{Adj}(G,X)$ is the set of vertices
adjacent to \textit{X}. If all of the
edges in a graph \textit{G} are directed edges, then \textit{G} is a
\textit{directed graph}. A~\textit{path} between $X_1$ and $X_n$ in \textit{G} is an
ordered sequence of vertices $\langle X_1,\ldots,X_n\rangle$ such
that for $1 < i
\leq n$, $X_{i-1}$ and $X_i$ are adjacent in \textit{G}. A path between
$X_1$ and $X_n$ in \textit{G} is a \textit{directed path} if for $1 <
i \leq n$, the edge between $X_{i-1}$ and $X_i$ is a directed edge
from $X_{i-1}$ to $X_i$. A~path is \textit{acyclic} if no vertex
occurs on
the path twice. A~directed graph is \textit{acyclic} (DAG) if all directed
paths are acyclic. \textit{X} is a \textit{parent} of \textit{Y} and \textit{Y} is a
\textit{child} of \textit{X} if there is an edge $X \rightarrow Y$. $
\langle X,Y,Z\rangle$ is
a \textit{triangle} in \textit{G} if \textit{X} is adjacent to \textit{Y} and \textit{Z}, and \textit{Y} is adjacent to \textit{Z}.

Suppose \textit{G} is a graph. \textbf{Parents}(\textit{G}, \textit{X}) is the set
of parents of \textit{X} in \textit{G}. \textit{X} is an \textit{ancestor} of \textit{Y} (and \textit{Y} is a \textit{descendant} of \textit{X}) if there is a
directed path from \textit{X} to \textit{Y}. A~subset of \textbf{V}
is \textit{ancestral}, if it is closed under the ancestor relation. A
triple of vertices $\langle X,Y,Z\rangle$ is \textit{unshielded} if \textit{X} is adjacent to \textit{Y} and \textit{Y} is adjacent to \textit{Z}, but \textit{X} is not adjacent to \textit{Z}. A triple of vertices $\langle
X,Y,Z\rangle$ is a
\textit{collider} if there are edges $X \rightarrow Y
\leftarrow Z$. A triple of vertices $\langle X,Y,Z\rangle$ is a
\textit{noncollider} if \textit{X} is adjacent to \textit{Y} and \textit{Y} is adjacent to \textit{Z}, but it is not a collider.

A probability distribution \textit{P} over a set of variables \textbf{V}
satisfies the (\textit{local directed}) \textit{Markov condition} for a DAG
\textit{G} if and only if each variable \textit{V} in \textbf{V} is independent of
the set of
variables that are neither parents nor descendants of \textit{V} in \textit{G}, conditional on the parents of \textit{V} in \textit{G}. A \textit{Bayesian
network} is an ordered pair $\langle P, G\rangle$ where $\it P$
satisfies the local directed Markov condition for \textit{G}. If $M =
\langle P, G\rangle$, $P_M$ denotes \textit{P} and $G_M$ denotes
\textit{G}. Two DAGs $G_1$ and $G_2$ over the same set of variables
\textbf{V} are said to be \textit{Markov equivalent} if all of the
conditional independence relations entailed by satisfying the local
directed Markov condition for $G_1$ are also entailed by satisfying
the local directed Markov condition for $G_2$, and vice versa. A
useful characterization of Markov equivalence between DAGs is that two
DAGs are Markov equivalent if and only if they have the same
adjacencies and the same unshielded colliders \citepp{VerPea90}.
A \textit{Markov equivalence class M} is a set of DAGs that contains all
DAGs that are Markov equivalent to each other. A Markov equivalence
class \textit{M} can be represented by a graph called a \textit{pattern}; a~pattern \textit{O} is a graph such that (i) if $X \rightarrow Y$ in
every DAG in \textit{M}, then $X \rightarrow Y$ in \textit{O}; and (ii) if
$X \rightarrow Y$ in some DAG in \textit{M} and $Y \rightarrow X $ in
some other DAG in \textit{M}, then \textit{X} --- \textit{Y} in \textit{O}. In that
case \textit{O} is said to \textit{represent} \textit{M} and each DAG in \textit{M}.

If \textbf{X} is independent of \textbf{Y} conditional on \textbf{Z}, we write $
I(\mathbf{X},\mathbf{Y}|\mathbf{Z})$, or if \textit{X}, \textit{Y}, and \textit{Z} are
individual variables $I(X,Y|Z)$. In a DAG \textit{G}, a vertex \textit{A}
is \textit{active} on an acyclic path \textit{U} between \textit{X} and \textit{Y}
conditional on set \textbf{Z} of vertices (not containing \textit{X} or \textit{Y}) if $A = X$ or $A = Y$, or \textit{A} is a noncollider on \textit{U}
and not in \textbf{Z}, or \textit{A} is a collider on \textit{U} that is in \textbf{Z} or has a descendant in \textbf{Z}. An acyclic path \textit{U} is
\textit{active} conditional on a set \textbf{Z} of vertices if every vertex on the
path is active relative to \textbf{Z}. If $X \neq Y$ and \textbf{Z} does not
contain \textit{X} or \textit{Y}, \textit{X} is \textit{d-separated} from \textit{Y}
conditional on \textbf{Z} if there is no active acyclic path between \textit{X} and \textit{Y} conditional on \textbf{Z}; otherwise \textit{X} and \textit{Y}
are \textit{d-connected} conditional on \textbf{Z}. For three disjoint sets
\textbf{X}, \textbf{Y} and \textbf{Z}, \textbf{X} is \textit{d-separated} from \textbf{Y}
conditional on \textbf{Z} if there is no acyclic active path between any
member of \textbf{X} and any member of \textbf{Y} conditional on \textbf{Z};
otherwise \textbf{X} and \textbf{Y} are \textit{d-connected} conditional on \textbf{Z}. If \textbf{X} is
d-separated from \textbf{Y} conditional on \textbf{Z} in
DAG \textit{G}, then $I(\mathbf{X},\mathbf{Y}|\mathbf{Z})$ in every probability
distribution that satisfies the local directed Markov condition for
\textit{G} \citepp{Pea88}. Any conditional independence relation that holds
in every distribution that satisfies the local directed Markov
condition for DAG \textit{G} is \textit{entailed by \textit{G}}. Note, however,
that in some distributions that satisfy the local directed Markov
condition for \textit{G}, some conditional independence relation $I(\mathbf{X},\mathbf{Y}|\mathbf{Z})$ may hold even if \textbf{X} is not d-separated from
\textbf{Y} conditional on \textbf{Z} in \textit{G}; such distributions are said
to be \textit{unfaithful to G}.

There are a number of different parameterizations of a DAG \textit{G},
which map \textit{G} onto distributions that satisfy the local directed
Markov condition for \textit{G}. One common parameterization is a
recursive linear Gaussian structural equation model. A recursive linear
Gaussian structural equation model is an ordered triple $\langle G,
\mathit{Eq}, \Sigma\rangle$, where \textit{G} is a DAG over a set of vertices $
X_1,\ldots,X_n$, \textit{Eq} is a set of equations, one for each $X_i$
such that
\[
{X}_{i}=\sum_{{{X}_{j}}\in\mathbf
{Parents}(G,{{X}_{i}})}{b}_{j,i}{X}_{j}+{
\varepsilon}_{i},
\]
where the $b_{j,i}$ are real constants known as the structural
coefficients, and the $\varepsilon_{i}$ are multivariate Gaussian
that are jointly independent of each other with covariance matrix $
\Sigma$. The $ \varepsilon_{i}$ are referred to as ``error terms.''
In vector notation, where \textbf{X} is the vector of $X_1,\ldots,X_n$,
\textbf{B} is the matrix of structural coefficients, and $\bolds
{\varepsilon}$ is the vector of error terms,
\[
\mathbf{X}=\mathbf{BX}+\bolds{\varepsilon}.
\]
The covariance matrix $\Sigma$ over the error terms, together with
the structural equations, determine a distribution over the variables
in \textbf{X}, which satisfies the local directed Markov condition for
\textit{G}. Hence, the DAG in a recursive linear Gaussian structural
equation model \textit{M} together with the probability distribution
generated by the equations and the covariance matrix over the error
terms form a Bayesian network. Because the joint distribution over the
nonerror terms of a linear Gaussian structural equation model is
multivariate Gaussian, \textit{X} is independent of \textit{Y} conditional on
\textbf{Z} in $P_M$ if and only if
$\rho_M(X,Y|\mathbf{Z}) = 0$, where $\rho_M(X,Y|\mathbf{Z})$ denotes the
conditional or partial correlation between \textit{X} and \textit{Y}
conditional on \textbf{Z} according to $P_M$. Let $e_M(X \rightarrow
Z)$ denote the structural coefficient of the $X \rightarrow Z$ edge
in $G_M$. If there is no edge $X \rightarrow Z$ in $G_M$, then $
e_M(X \rightarrow Z) = 0$. If \textit{X} and \textit{Z} are adjacent in $
G_M$, then $e_M(X \mbox{ --- } Z) = e_M(X \rightarrow Z)$ if there is an $X
\rightarrow Z$ edge in $G_M$, and otherwise $e_M(X \mbox{ --- } Z) = e_M(Z
\rightarrow X)$.

There is a \textit{causal interpretation} of recursive linear Gaussian
structural equation models, in which setting (as in an experiment, as
opposed to observing) the value of $X_i$ to the fixed value \textit{x}
is represented by replacing the structural equation for $X_i$ with
the equation $X_i = x$. Under the causal interpretation, a recursive
linear structural equation model is a \textit{causal model}, the DAG $
G_M$ is a \textit{causal DAG}, and the pattern that represents $G_M$ is
a \textit{causal pattern}. A causal model with a set of variables \textbf{V}
is \textit{causally sufficient} when every common direct cause of any two
variables in \textbf{V} is also in \textbf{V}. Informally, under a causal
interpretation, an edge $X \rightarrow Y$ in $G_M$ represents that
\textit{X} is a direct cause of \textit{Y} relative to \textbf{V}. A causal
model of a population is true when the model correctly predicts the
results of all possible settings of any subset of the variables \citepp{Pea00}.

There are two assumptions made about the relationship between the
causal DAG and the population probability distribution that play a key
role in causal inference from observational data. A discussion of the
implications of these assumptions, arguments for them, and a discussion
of conditions when they should not be assumed are given in \citeauthor{SpiGlySch93}
(\citeyear{SpiGlySch93}), pages~32--42. In this paper, we will consider only those cases
where the causal relations in a given population can be represented by
a model whose graph is a DAG.

\textit{Causal Markov assumption \textup{(}CMA\textup{)}}. If the true causal model \textit{M}
of a population is causally sufficient, every variable in \textbf{V} is
independent of the variables that are neither its parents nor
descendants in $G_M$ conditional on its parents in $G_M$.

\textit{Causal Faithfulness assumption \textup{(}CFA\textup{)}}. Every conditional
independence relation that holds in the population probability
distribution is entailed by the true causal DAG of the population.

The Causal Markov and Causal Faithfulness assumptions together entail
that \textbf{X} is independent of \textbf{Y} conditional on \textbf{Z} in the
population if and only if \textbf{X} is d-separated from \textbf{Y}
conditional on \textbf{Z} in the true causal graph.

\section{Weakening the Causal Faithfulness Assumption}\label{sec3}
A number of algorithms for causal estimation have been proposed that
rely on the assumption of the causal sufficiency of the observed
variables, the Causal Markov assumption and the Causal Faithfulness
assumption. The \textit{SGS} algorithm (\citeauthor{SpiGlySch93}, \citeyear{SpiGlySch93}, pa\-ge~82), for
example, is a Fisher consistent estimator of causal patterns under
these assumptions. (This, together with a uniformly consistent test of
conditional independence, entails that the \textit{SGS} algorithm is a
pointwise consistent estimator of causal patterns.)

In this section we explore ways to weaken the Causal Faithfulness
assumption that still allow pointwise consistent estimation of
(features of) causal structure, and we illustrate the ideas by going
through a sequence of generalizations of the population version of the
\textit{SGS} algorithm. None of the results in this section depend upon
assuming Gaussianity or linearity. The basic idea is that although the
Causal Faithfulness assumption is not fully testable (without knowing
the true causal structure), it has testable components given the Causal
Markov assumption. Under the Causal Markov assumption, the Causal
Faithfulness assumption entails that the probability distribution
admits a perfect DAG representation, that is, a DAG that entails all
and only those conditional independence relations true of the
distribution. Whether there is such a DAG depends only on the
distribution, and so is, in theory, testable. In principle, then, one
may adopt a weaker-than-faithfulness assumption and test (rather than
assume) the testable part of the faithfulness condition.

The \textit{SGS} algorithm takes an oracle of conditional independence as
input, and outputs a graph on the given set of variables with both
directed edges and undirected edges.

\textit{SGS algorithm}.
\begin{enumerate}[S2.]

\item[S1.] Form the complete undirected graph \textit{H} on the given set
of variables \textbf{V}.

\item[S2.] For each pair of variables \textit{X} and \textit{Y} in \textbf{V},
search for a subset \textbf{S} of $\mathbf{V}\setminus\{X, Y\}$ such that
\textit{X} and \textit{Y} are independent conditional on \textbf{S}. Remove the
edge between \textit{X} and \textit{Y} in \textit{H} if and only if such a set
is found.

\item[S3.] Let \textit{K} be the graph resulting from S2. For each
unshielded triple $\langle X, Y, Z\rangle$ (i.e., \textit{X} and \textit{Y} are adjacent, \textit{Y} and \textit{Z} are adjacent, but \textit{X} and
\textit{Z} are not adjacent),
\begin{enumerate}[(ii)]
\item[(i)] If \textit{X} and \textit{Z} are not independent conditional on
any subset of $\mathbf{V}\setminus\{X, Z\}$ that contains \textit{Y}, then
orient the triple as a collider: $X \rightarrow Y \leftarrow Z$.
\item[(ii)] If \textit{X} and \textit{Z} are not independent conditional on
any subset of $\mathbf{V}\setminus\{X, Z\}$ that does not contain \textit{Y}, then mark the triple as a noncollider (i.e., not $X \rightarrow Y
\leftarrow Z$).
\end{enumerate}
\item[S4.] Execute the following orientation rules until none of them applies:
\begin{enumerate}[(iii)]
\item[(i)] If $X \rightarrow Y\mbox{ --- }Z$, and the triple $\langle X, Y,
Z\rangle$ is marked as a noncollider, then orient $Y\mbox{ --- }Z$ as $Y
\rightarrow Z$.
\item[(ii)] If $X \rightarrow Y \rightarrow Z$ and $X\mbox{ --- }Z$, then
orient $X\mbox{ --- }Z$ as $X \rightarrow Z$.
\item[(iii)] If $X \rightarrow Y \leftarrow Z$, another triple $
\langle X, W, Z\rangle$ is marked as a noncollider, and $W\mbox{ --- }Y$,
then orient $W\mbox{ --- }Y$ as $W \rightarrow Y$. (This rule was not in
the original \textit{SGS} or \textit{PC} algorithm, but added by \citeauthor{Mee95}, \citeyear{Mee95}.)
\end{enumerate}
\end{enumerate}

Assuming the oracle of conditional independence is perfectly reliable
(which we will do throughout this section), the \textit{SGS} algorithm is
correct under the Causal Markov and Faithfulness assumptions, in the
sense that its output is the pattern that represents the~Markov
equivalence class containing the true causal DAG (\citeauthor{SpiGlySch93}, \citeyear{SpiGlySch93},
page~82; \citeauthor{Mee95}, \citeyear{Mee95}).

The correctness of \textit{SGS} follows from the following three
properties of d-separation (\cite{SpiGlySch93}):
\begin{longlist}[1.]
\item[1.]\textit{X} is adjacent to \textit{Y} in DAG \textit{G} if and only if \textit{X} is not d-separated from \textit{Y} conditional on any subset of the
other variables in \textit{G}.
\item[2.] If $\langle X, Y, Z\rangle$ is an unshielded collider in DAG
\textit{G}, then \textit{X} is not d-separated from \textit{Z} conditional on
any subset of the other variables in \textit{G} that contains \textit{Y}.
\item[3.] If $\langle X, Y, Z\rangle$ is an unshielded noncollider in
DAG \textit{G}, then \textit{X} is not d-separated from \textit{Z} conditional
on any subset of the other variables in \textit{G} that does not contain~\textit{Y}.
\end{longlist}

We shall not reproduce the full proof here, but a few points are worth
stressing. First, S2 is the step of inferring adjacencies and
nonadjacencies. The inferred adjacencies, represented by the remaining
edges in the graph resulting from S2, are correct because of the Causal
Markov assumption alone: every DAG Markov to the given oracle must
contain at least these adjacencies. On the other hand, the inferred
nonadjacencies (via removal of edges) are correct because of the Causal
Faithfulness assumption, or, more precisely, because of the following
consequence of the Causal Faithfulness assumption, which we, following
\citet{RamZhaSpi06}, will refer to as Adjacency-Faithfulness.

\textit{Adjacency-Faithfulness assumption}.
Given a set of variables \textbf{V} whose true causal DAG is \textit{G}, if two variables \textit{X}, \textit{Y}
are adjacent in \textit{G}, then they are not independent conditional on
any subset of $\mathbf{V}\setminus\{X, Y\}$.

Under the Adjacency-Faithfulness assumption, any edge removed in S2 is
correctly removed, because any DAG with the adjacency violates the
Adjacency-Faithfulness assumption.

Second, the key step of inferring orientations is step S3, in which
unshielded colliders and noncolliders are inferred. Given that the
adjacencies and nonadjacencies are all correct, the clauses (i) and
(ii) in step S3, as formulated here, are justified by the Causal Markov
assumption alone. Take clause (i), for example. If the unshielded triple
$\langle X, Y, Z\rangle$ is not a collider in the true causal DAG,
then the Causal Markov assumption entails that \textit{X} and \textit{Z} are
independent conditional on some set that contains \textit{Y}. That is why
clause (i) is sound. A~similar argument shows that clause (ii) is
sound. This does not mean, however, that the Causal Faithfulness
assumption does not play any role in justifying S3. Notice that the
antecedent of (i) and that of (ii) do not exhaust the logical
possibilities. They leave out the possibility that \textit{X} and \textit{Z}
are independent conditional on some set that contains \textit{Y} and
independent conditional on some set that does not contain \textit{Y}. This
omission is justified by the Causal Faithfulness assumption, or, more
precisely, by the following consequence of the Causal Faithfulness
assumption (\cite{RamZhaSpi06}):

\textit{Orientation-Faithfulness assumption.} Given a set of variables \textbf{V}
whose true causal DAG is \textit{G}, let $\langle X, Y, Z\rangle$ be
any unshielded triple in \textit{G}:
\begin{longlist}[1.]
\item[1.] If $X \rightarrow Y \leftarrow Z$, then \textit{X} and \textit{Z} are
not independent conditional on any subset of $\mathbf{V}\setminus\{X, Z\}
$ that contains~\textit{Y};
\item[2.] Otherwise, \textit{X} and \textit{Z} are not independent conditional on
any subset of $\mathbf{V}\setminus\{X, Z\}$ that does not contain~\textit{Y}.
\end{longlist}

Obviously, the possibility left out by S3 is indeed ruled out by the
Orientation-Faithfulness assumption.

The Orientation-Faithfulness assumption, if true, justifies a much
simpler and more efficient step than S3: for every unshielded triple $
\langle X, Y, Z\rangle$, we need check only the set found in S2 that
renders \textit{X} and \textit{Z} independent; the triple is a collider if
and only if the set does not contain \textit{Y}. This simplification is
used in the \textit{PC} algorithm, a well-known, more computationally
efficient rendition of the \textit{SGS} procedure (\citeauthor{SpiGlySch93},
\citeyear{SpiGlySch93},
pages~84--85). Moreover, the Adjacency-Faithfulness condition also
justifies a couple of measures to improve the efficiency of S2, used by
the \textit{PC} algorithm. Here we are concerned with showing how the
basic \textit{SGS} procedure may be modified to be correct under
increasingly weaker assumptions of faithfulness, so we will not go into
the details of the optimization measures in the \textit{PC} algorithm.
Whether these or similar measures are available to the modified
algorithms we introduce below is an important question to be addressed
in future work.

Let us start with the modification proposed by \citet{RamZhaSpi06},
who observed that assuming the Causal Markov and Adjacency-Faithfulness
assumptions are true, any failure of the Orientation-Faithfulness
assumption is \textit{detectable}, in the sense that the probability
distribution in question is not both Markov and Faithful to any DAG
\citepp{ZhaSpi08}. In our formulation of the \textit{SGS}
algorithm, it is easy to see how failures of Orientation-Faithfulness
can be detected. As already mentioned, the role of the
Orientation-Faithfulness assumption in justifying the \textit{SGS}
algorithm is to guarantee that at the step S3, either the antecedent of
(i) or that of (ii) will obtain. Therefore, if it turns out that for
some unshielded triple neither antecedent is satisfied, the
Orientation-Faithfulness assumption is detected to be false for that triple.

This suggests a simple modification to S3 in the \textit{SGS} algorithm.

S3*. Let \textit{K} be the undirected graph resulting from S2. For each
unshielded triple $\langle X, Y, Z\rangle$,
\begin{longlist}[(iii)]
\item[(i)] If \textit{X} and \textit{Z} are not independent conditional on
any subset of $\mathbf{V}\setminus\{X, Z\}$ that contains \textit{Y}, then
orient the triple as a collider: $X \rightarrow Y \leftarrow Z$.
\item[(ii)] If \textit{X} and \textit{Z} are not independent conditional on
any subset of $\mathbf{V}\setminus\{X, Z\}$ that does not contain \textit{Y},
then mark the triple as a noncollider.
\item[(iii)] Otherwise, mark the triple as ambiguous (or unfaithful).
\end{longlist}

\citet{RamZhaSpi06} applied essentially this modification to the \textit{PC} algorithm and called the resulting algorithm the \textit{Conservative
PC \textup{(}CPC\textup{)}} algorithm. (Their results show that the main optimization
measures used in the \textit{PC} algorithm still apply to this
generalization of \textit{SGS} because the Adjacency-Faithfulness
condition is still assumed.) We will thus call the algorithm that
results from replacing S3 with S3* the \textit{Conservative SGS \textup{(}CSGS\textup{)}} algorithm.

\begin{figure}

\includegraphics{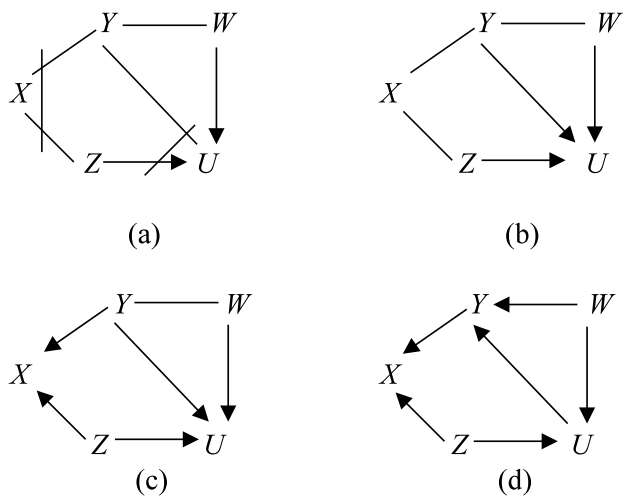}

\caption{\textup{(a)} is a sample output of the \textit{CSGS} algorithm. The
ambiguous (or unfaithful) unshielded triples are marked by straight
lines crossing the two edges. There is no explicit mark for
noncolliders, with the understanding that all and only unshielded
triples that are not oriented as colliders or marked as ambiguous are
(implicitly) marked noncolliders. \textup{(b)--(d)} are the three patterns
represented by \textup{(a)}.}

\label{penG}
\end{figure}

It is straightforward to prove that the \textit{CSGS} algorithm is correct
under the Causal Markov and the Adjacency-Faithfulness assumptions
alone, in the sense that if the Causal Markov and
Adjacency-Faithfulness assumptions are true and if the oracle of
conditional independence is perfectly reliable, then every adjacency,
nonadjacency, orientation and marked noncollider in the output of the
\textit{CSGS} are correct. As pointed out in \citet{RamZhaSpi06}, the
output of the \textit{CSGS} can be understood as an \textit{extended pattern}
that represents a set of patterns. For example, a sample output used in
\citet{RamZhaSpi06} is given in Figure~\ref{penG}(a). There are two ambiguous
unshielded triples in the output: $\langle Y, X, Z\rangle$ and $
\langle Z, U, Y\rangle$, which are marked by crossing straight lines.
Note that there is no explicit mark for noncolliders, with the
understanding that all and only unshielded triples that are not
oriented as colliders or marked as ambiguous are (implicitly) marked
noncolliders. Figure~\ref{penG}(a) represents a set of three patterns, depicted
in Figure~\ref{penG}(b)--(d). Each pattern results from some disambiguation of
the ambiguous triples in Figure~\ref{penG}(a). The pattern in Figure~\ref{penG}(b), for
example, results from taking the triple $\langle Y, X, Z\rangle$ as
a noncollider and taking the triple $\langle Z, U, Y\rangle$ as a
collider. Note that not every disambiguation results in a pattern.
Taking both ambiguous triples as noncolliders would force a directed
cycle: $Z \rightarrow U \rightarrow Y \rightarrow X \rightarrow Z$,
and so would not lead to a pattern. That is why there are only three
instead of four patterns in the set represented by Figure~\ref{penG}(a).

It is easy to see that when the Orientation-Faithfulness assumption
happens to hold, the \textit{CSGS} output will be a single pattern (i.e.,
without ambiguous triples), which is the same as the \textit{SGS }output.
In other words, \textit{CSGS} is as informative as \textit{SGS} when the
stronger assumption needed for the output of the latter to be
guaranteed to be correct happens to be true.

The Adjacency-Faithfulness assumption may be further weakened. In an
earlier paper (Zhang and Spirtes, \citeyear{ZhaSpi08}), we showed that some violations
of the Adjacency-Faithfulness assumption are also detectable, and we
specified some conditions weaker than the Adjacency-Faithfulness
assumption under which any violation of Faithfulness (and so any
violation of Adjacency-Faithfulness) is detectable. One of the weaker
conditions is known as the Causal Minimality assumption (\citeauthor{SpiGlySch93}, \citeyear{SpiGlySch93}, page~31), which states that the true causal DAG is a minimal
DAG that satisfies the Markov condition with the true probability
distribution, minimal in the sense that no proper subgraph satisfies
the Markov condition. This condition is a consequence of the
Adjacency-Faithfulness assumption. If the Adjacency-Faithfulness
assumption is true, then no edge can be taken away from the true causal
DAG without violating the Markov condition.

The other weaker condition is named Triangle-Faithfulness:

\textit{Triangle-Faithfulness assumption}. Suppose the true causal DAG of
\textbf{V} is \textit{G}. Let \textit{X}, \textit{Y}, \textit{Z} be any three
variables that form a triangle in \textit{G} (i.e., each pair of vertices
is adjacent):
\begin{longlist}[1.]
\item[1.] If \textit{Y} is a noncollider on the path $\langle X, Y, Z\rangle
$, then \textit{X} and \textit{Z} are not independent conditional on any
subset of $\mathbf{V}\setminus\{X, Z\}$ that does not contain \textit{Y};
\item[2.] If \textit{Y} is a collider on the path $\langle X, Y, Z\rangle$,
then \textit{X} and \textit{Z} are not independent conditional on any subset
of $\mathbf{V}\setminus\{X, Z\}$ that contains \textit{Y}.
\end{longlist}

Clearly, the Adjacency-Faithfulness assumption entails the
Triangle-Faithfulness assumption, and the latter, intuitively, is much
weaker. Our result in \citet{ZhaSpi08} is that given the Causal
Markov, Minimality and Triangle-Faithfulness assumptions, any violation
of faithfulness is detectable. But we did not propose any algorithm
that is provably correct under the Markov, Minimality and
Triangle-Faithfulness assumptions.

What need we modify in the \textit{SGS} algorithm if all we can assume are
the Markov, Minimality and Triangle-Faithfulness assumptions? In the
step S2, the inferred adjacencies are still correct, which, as already
mentioned, is guaranteed by the Causal Markov assumption alone. The
inferred nonadjacencies, however, are not necessarily correct, because
the Adjacency-Faithfulness assumption might fail. So the first
modification we need make is to acknowledge that the nonadjacencies
resulting from S2 are only ``apparent'' but not ``definite'': there
might still be an edge between two variables even though the edge
between them was removed in S2 because a screen-off set was found.

Since we do not assume the Orientation-Faithful\-ness assumption,
obviously we need at least modify S3 into S3*. A further worry is that
the unshielded triples resulting from S2 are only ``apparent'': they
might be shielded in the true causal DAG but appear to be unshielded
due to a failure of Adjacency-Faithfulness. Fortunately, this
possibility does not affect the soundness of S3*. Take clause (i) for
example. For an apparently unshielded triple $\langle X, Y, Z\rangle$,
either \textit{X} and \textit{Z} are really nonadjacent in the true DAG or
they are adjacent. In the former case, clause (i) is sound by the
Markov assumption. In the latter case, clause (i) is still sound by the
Triangle-Faithfulness assumption. A similar argument shows that clause
(ii) is also sound. So S3* is still sound. Moreover, clause (iii) can
now play a bigger role than simply conceding ignorance or ambiguity. If
the antecedent of clause (iii) is satisfied, then one can infer that
\textit{X} and \textit{Z} are really nonadjacent, for otherwise the
Triangle-Faithfulness assumption would be violated no matter whether $
\langle X, Y, Z\rangle$ is a collider or not.

The soundness of S4 is obviously not affected. Therefore, if we only
assume the Causal Markov, Minimality and Triangle-Faithfulness
assumptions, the \textit{CSGS} algorithm is still correct if we take the
nonadjacencies in its output as uninformative (except for those
warranted by S3*).

The question now is whether we can somehow test the
Adjacency-Faithfulness assumption in the procedure and confirm the
nonadjacencies when the test returns affirmative. The following lemma
gives a sufficient condition for verifying the Adjacency-Faithfulness
assumption and hence the nonadjacencies in the \textit{CSGS} output.
(Recall that the \textit{CSGS} output in general represents a set of
patterns, and each pattern represents a set of Markov equivalent DAGs.)
A pattern \textit{O} is \textit{Markov to an oracle} when for every DAG
represented by \textit{O}, each vertex is independent of the set of
variables that are neither descendants nor parents in the DAG
conditional on the parents in the DAG according to the oracle.

\begin{lemma}\label{le1}
Suppose the Causal Markov, Minimality and Triangle-Faithfulness
assumptions are true, and \textit{E} is the output of \textit{CSGS} given a
perfectly reliable oracle of conditional independence. If every pattern
in the set represented by \textit{E} is Markov to the oracle, then the
true causal DAG has exactly those adjacencies present in \textit{E}.
\end{lemma}
\begin{pf} As we already pointed out, the true causal DAG, $G_T$,
must have at least the adjacencies in \textit{E} (in order to satisfy the
Causal Markov assumption), and must have the colliders and noncolliders
in \textit{E} (in order to satisfy the Causal Markov and the
Triangle-Faithfulness assumptions). Now suppose every pattern in the
set represented by \textit{E} is Markov to the oracle, and suppose, for
the sake of contradiction, that $G_T$ has still more adjacencies. Let
\textit{G} be the proper subgraph of $G_T$ with just the adjacencies in
\textit{E}. Then every unshielded collider and every unshielded
noncollider in \textit{E} are also present in \textit{G}, and other
unshielded triples in \textit{G}, if any, are ambiguous in \textit{E}. Thus,
the pattern that represents the Markov equivalence class of \textit{G} is
in the set represented by \textit{E}. It follows that \textit{G} is Markov to
the oracle, which shows that $G_T$ is not a minimal graph that is
Markov to the oracle. This contradicts the Causal Minimality
assumption. Therefore, $G_T$ has exactly the adjacencies present in
\textit{E}.
\end{pf}

So we have the following \textit{Very Conservative SGS} (\textit{VCSGS}):

\textit{VCSGS algorithm}.
\begin{enumerate}[V1.]
\item[V1.] Form the complete undirected graph \textit{H} on the given set
of variables \textbf{V}.
\item[V2.] For each pair of variables \textit{X} and \textit{Y} in \textbf{V},
search for a subset \textbf{S} of $\mathbf{V}\setminus\{X, Y\}$ such that
\textit{X} and \textit{Y} are independent conditional on \textbf{S}. Remove the
edge between \textit{X} and \textit{Y} in \textit{H} and mark the pair $\langle
X, Y\rangle$ as ``apparently nonadjacent,'' if and only if such a set
is found.
\item[V3.] Let \textit{K} be the graph resulting from V2. For each
apparently unshielded triple $\langle X, Y, Z\rangle$ (i.e., \textit{X}
and \textit{Y} are adjacent, \textit{Y} and \textit{Z} are adjacent, but \textit{X}
and \textit{Z} are apparently nonadjacent),
\begin{enumerate}[(iii)]
\item[(i)] If \textit{X} and \textit{Z} are not independent conditional on
any subset of $\mathbf{V}\setminus\{X, Z\}$ that contains \textit{Y}, then
orient the triple as a collider: $X \rightarrow Y \leftarrow Z$.
\item[(ii)] If \textit{X} and \textit{Z} are not independent conditional on
any subset of $\mathbf{V}\setminus\{X, Z\}$ that does not contain \textit{Y}, then mark the triple as a noncollider.
\item[(iii)] Otherwise, mark the triple as ambiguous (or unfaithful),
and mark the pair $\langle X, Z\rangle$ as ``definitely nonadjacent.''
\end{enumerate}
\item[V4.] Execute the same orientation rules as in S4, until none of
them applies.
\item[V5.] Let \textit{M} be the graph resulting from V4. For each
consistent disambiguation of the ambiguous triples in \textit{M} (i.e.,
each disambiguation that leads to a pattern), test whether the
resulting pattern satisfies the Markov condition. If every pattern
does, then mark all the ``apparently nonadjacent'' pairs as
``definitely nonadjacent.''
\end{enumerate}

[An obvious way to test the Markov condition in V5 on a given pattern
is to extend the pattern to a DAG and test the local Markov condition.
That is, we need to test, for each variable \textit{X}, whether \textit{X} is
independent of the variables that are neither its descendants nor its
parents conditional on its parents. In linear Gaussian models, this can
be done by regressing \textit{X} on its nondescendants and testing whether
the regression coefficients are zero for its nonparents. More
generally, assuming composition, we need only run a conditional
independence test for each nonadjacent pair, and, thus, in the worst
case the number of conditional independence tests is $O(n^2)$, where
\textit{n} is the number of vertices. The number of patterns to be tested
in V5 is $O(2^a)$, where \textit{a} is the number of ambiguous
unshielded triples.]

\begin{figure}[b]

\includegraphics{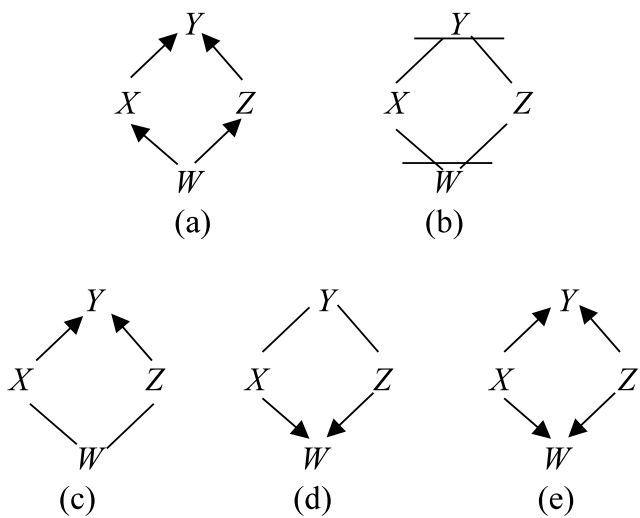}

  \caption{An example in which the test in step V5 of VCSGS does not confirm
  the nonadjacencies even though the nonadjacencies are correct.}\label{fig2}
\end{figure}

As we already explained, steps V1--V4 are sound under the Causal Markov,
Minimality and Triangle-Faithfulness assumptions. Lemma \ref{le1} shows that V5
is also sound. Hence, the \textit{VCSGS} algorithm is correct under the
Causal Markov, Minimality and Triangle-Faithfulness assumptions, in the
sense that given a perfectly reliable oracle of conditional
independence, all the adjacencies, definite nonadjacencies, directed
edges and marked noncolliders are correct. Moreover, when the Causal
Faithfulness assumption happens to hold, the \textit{CSGS} output will be
a single pattern and this single pattern will satisfy the Markov
condition; hence, the \textit{VCSGS} algorithm will return a single
pattern with full information about nonadjacencies. Therefore, \textit{VCSGS} is also as informative as \textit{SGS} when the Causal Faithfulness
assumption happens to be true.

One might think (or hope) that the \textit{VCSGS} algorithm is as
informative as the \textit{CSGS} algorithm when Adjacency-Faithfulness
(but not Orientation-Faithfulness) happens to hold. Unfortunately this
is not true in general because the sufficient condition given in Lemma
\ref{le1} (and checked in V5) is not necessary for the Adjacency-Faithfulness
assumption.

To illustrate, consider the following example. Suppose the true causal
DAG is the one given in Figure~\ref{fig2}(a). Suppose the causal Markov
assumption and the Adjacency-Faithfulness assumption are satisfied. And
suppose that, besides the conditional independence relations entailed
by the graph, the true distribution features one and only one extra
conditional independence: $I(X, Z | Y)$, due, for example, to some
sort of balancing-out of the path $\langle X, Y, Z\rangle$ (active
conditional on $\{\textit{Y}\}$) and the path $\langle X, W, Z\rangle
$ (active conditional on $\{\textit{Y}\}$). This violates the
Orientation-Faithfulness assumption. The \textit{CSGS} output will thus be
the graph in Figure~\ref{fig2}(b), in which both the triple $\langle X, Y,
Z\rangle$ and the triple $\langle X, W, Z\rangle$ are ambiguous.
This output represents a set of three patterns, as shown in Figure~\ref{fig2}(c)--(e). (Again, the two ambiguous triples cannot be noncolliders at
the same time.) However, only the patterns in Figure~\ref{fig2}(c) and \ref{fig2}(d) satisfy the
Markov condition. The pattern in Figure~\ref{fig2}(e) violates the Markov condition
because it entails that $I(X, Z | \varnothing)$, which is not true.

For this example, then, the \textit{VCSGS} will not return the full
information of nonadjacencies, even though the Adjacency-Faithfulness
assumption is true.

In light of this example, it is natural to consider the following
variant of step V5 in \textit{VCSGS}:

V5*. Let \textit{M} be the graph resulting from V4. If \textit{some}
disambiguation of the ambiguous triples in \textit{M} leads to a pattern
that satisfies the Markov condition, then mark all remaining
``apparently nonadjacent'' pairs as ``definitely nonadjacent.''

We suspect that V5* is also sound under the Causal Markov, Minimality
and Triangle-Faithfulness assumptions, but we have not found a proof.
In other words, we conjecture that the sufficient condition presented
in Lemma \ref{le1} can be weakened to that \textit{some} pattern in the set
represented by the \textit{CSGS} output satisfies the Markov condition.
(This conjecture is a consequence of the following plausible
conjecture: Suppose a DAG \textit{G} and a probability distribution \textit{P} satisfy the Markov, Minimality and Triangle-Faithfulness conditions.
Then no DAG with strictly fewer adjacencies than in \textit{G} is Markov
to \textit{P}. We thank an anonymous referee for making the point and the
conjecture.) Note that if the Adjacency-Faithfulness assumption happens
to hold, then at least one pattern (i.e., the pattern representing the
true causal DAG) satisfies the Markov condition. Therefore, if our
conjecture is true, we can replace V5 with V5* in the \textit{VCSGS}
algorithm, and the condition tested in V5* is both sufficient and
necessary for Adjacency-Faithfulness. The resulting algorithm will then
be as informative as the \textit{CSGS} algorithm whenever the
Adjacency-Faithfulness assumption happens to hold, and as informative
as the \textit{SGS} algorithm whenever both the Adjacency-Faithfulness
assumption and the Orientation-Faithfulness assumption happen to hold.

It is worth noting that if we adopt a natural, interventionist
conception of causation (e.g., \cite{Woo03}), the Causal Minimality
assumption is guaranteed to be true if the probability distribution is
positive \citepp{ZhaSpi11}. Since positivity is a property of
the probability distribution alone, we may also try to incorporate a
test of positivity at the beginning of \textit{VCSGS}, and proceed only if
the test returns affirmative. We then need not assume the Causal
Minimality assumption in order to justify the procedure.

\section{Weakening the Strong Causal Faithfulness Assumption}\label{sec4}
In this section we consider sample versions of the \textit{CSGS} and \textit{VCSGS} algorithms, assuming Gaussianity and linearity, and prove some
positive results on uniform consistency, under a generalization and
strengthening of the Triangle-Faithfulness assumption, which we call
the \textit{k}-Triangle-Faithfulness assumption.

If a model \textit{M} does not satisfy the Causal Faithfulness assumption,
then \textit{M} contains a zero partial correlation $\rho_M(X,Y|\mathbf{W})$
even though the Causal Markov assumption does not entail that $\rho
_M(X,Y|\mathbf{W})$ is zero. If $\rho_M(X,Y|\mathbf{W}) = 0$ but is not
entailed to be zero for all values of the parameters, the parameters of
the model satisfy an algebraic constraint. A set of parameters that
satisfies such an algebraic constraint is a ``surface of
unfaithfulness'' in the parameter space that is of a lower dimension
than the full parameter space. Lying on such a surface of
unfaithfulness is of Lebesgue measure zero. For a Bayesian with a prior
probability over the parameter space that is absolutely continuous with
Lebesgue measure, the prior probability of unfaithfulness is zero.

However, in practice, the \textit{SGS} (or \textit{PC}) algorithm does not
have access to the population correlation coefficients. Instead it
performs statistical tests of whether a partial correlation is zero. If
$|\rho_M(X,Y|\mathbf{W})|$ is small enough, then with high probability a
statistical test of whether $\rho_M(X,Y|\mathbf{W})$ equals zero will not
reject the null hypothesis. If $\rho_M(X,Y|\mathbf{W})=0$ fails to be
rejected, this can lead to some edges that occur in the true causal DAG
not appearing in the output of \textit{SGS} and to errors in the
orientation of edges in the output of \textit{SGS}. (Such errors can also
lead to the output of the \textit{SGS} algorithm to fail to be a pattern,
either because it contains double-headed edges or undirected nonchordal
cycles.) \citet{Robetal03} showed that even if it is assumed that
there are no unfaithful models, there are always models so ``close to
unfaithful'' [i.e., with $\vert\rho_M(X,Y| \mathbf{W})\vert$ nonzero but small
enough that a statistical test will probably fail to reject the null
hypothesis] that there is no algorithm that is a uniformly consistent
estimator of the pattern of a causal model.

\citet{KalBuh07} showed that under a strengthened
version of the Causal Faithfulness assumption, the \textit{PC} algorithm
is a uniformly consistent estimator of the pattern that represents the
true causal DAG. Let \textit{n} be the sample size.
Their strengthened set of assumptions were as follows:

\begin{longlist}[(A1)]
\item[(A1)] The distribution $P_n$ is multivariate Gaussian and
faithful to the DAG $G_n$ for all \textit{n}.
\item[(A2)] The dimension $p_n = O(n^a)$ for some $0 \leq a < \infty
$.
\item[(A3)] The maximal number of neighbors in the DAG $G_n$ is
denoted by
\begin{eqnarray}
{{q}_{n}}=\mathop{\max}_{1\le j\le{{p}_{n}}} \bigl\llvert \operatorname{adj}(G,j) \bigr
\rrvert\nonumber \\
\eqntext{\mbox{with } {{q}_{n}}=O\bigl({{n}^{1-b}}\bigr)\mbox{ for some }0<b\le1.}
\end{eqnarray}
\item[(A4)] The partial correlations between $\mathbf{X}(i)$ and $\mathbf{X}(j)$ given $\{\mathbf{X}(r); r \in\mathbf{k}\}$ for some set $\mathbf{k}
\subseteq
\{1,\ldots,p_n\}\setminus\{i,j\}$ are denoted by $\rho_{n;i,j|\mathbf{k}}$. Their absolute values are bounded from below and above:
%
%
\begin{eqnarray}
\inf \bigl\{ {\llvert {{\rho_{i,j|{\mathbf{k}}}}} \rrvert ;i,j,{\mathbf {k}} {\mbox{
with }} {\rho_{i,j|{\mathbf{k}}}} \ne0} \bigr\} &\geq& {c_n},\nonumber\\
c_n^{ - 1}
&= &O\bigl({n^d}\bigr),\nonumber\\
 \eqntext{{\mbox{for some }}0 < d < b/2,}
\\
\mathop{\sup} _{n;i,j,{\mathbf{k}}} \llvert {{\rho _{i,j|{\mathbf{k}}}}} \rrvert& \leq& M
<
1,\nonumber\\
 \eqntext{{\mbox{where }}0 < b \leq1{\mbox{ is as in (A3).}}}
\end{eqnarray}
%
\end{longlist}

We will refer to the assumption that all nonzero partial correlations
are bounded below in absolute value by a number greater than zero [as in the first part
of (A4)] as the Strong Causal Faithfulness assumption. \citet{Uhl12} provide some reason to believe that unless $c_n$ is quite
small, the probability of violating Strong Causal Faithfulness
assumption is high, especially when the number of variables is large.
[This problem with assumption (A4) is somewhat mitigated by the fact
that the size of $c_n$ can decrease with increasing sample size. But
see \citet{Linetal12}, for an interesting analysis of the asymptotics
when $c_n$ approaches zero.]

It is difficult to see how a uniformly consistent estimator of a causal
pattern would be possible without assuming something like the Strong
Causal Faithfulness assumption. However, what we will show is that it
is possible to weaken the Strong Causal Faithfulness assumption in
several ways as long as the standard of success is not finding a
uniformly consistent estimator of the causal pattern, but is instead
finding a uniformly consistent estimator of (some of) the structural
coefficients in a pattern. The latter standard is compatible with
missing some edges that are in the true causal graph, as long as the
edges that have not been included in the output have sufficiently small
structural coefficients.

We propose to replace the faithfulness assumption in (A1), and the
Strong Faithfulness assumption with the following assumption, where $
e_M(X\mbox{ --- }Z)$, as we explained in Section~\ref{sec2}, denotes the structural
coefficient associated with the edge between \textit{X} and \textit{Z}.

\textit{k-Triangle-Faithfulness assumption}.
Given a set of variables \textbf{V}, suppose the true causal model over \textbf{V} is \textit{M} = $\langle
P,G\rangle$, where \textit{P} is a Gaussian distribution over \textbf{V},
and \textit{G} is a DAG with vertices \textbf{V}. For any three variables
\textit{X}, \textit{Y}, \textit{Z} that form a triangle in \textit{G} (i.e., each
pair of vertices is adjacent),
\begin{longlist}[1.]
\item[1.] If \textit{Y} is a noncollider on the path $\langle X, Y, Z\rangle
$, then $|\rho_M(X, Z|\mathbf{W})| \geq k \times|e_M(X\mbox{ --- }Z)|$ for
all $\mathbf{W} \subseteq\mathbf{V} $ that do not contain \textit{Y}; and
\item[2.] If \textit{Y} is a collider on the path $\langle X, Y, Z\rangle$,
then $|\rho_M(X, Z|\mathbf{W})| \geq k \times|e_M(X\mbox{ --- }Z)|$ for all $
\mathbf{W} \subseteq\mathbf{V} $ that do contain \textit{Y}.
\end{longlist}

As \textit{k} approaches 0, the \textit{k}-Triangle-Faithfulness assumption
approaches the Triangle-Faithfulness assumption. For (small) $k > 0$,
the \textit{k}-Triangle-Faithfulness assumption prohibits not only exact
cancellations of active paths in a triangle, but also \textit{almost}
cancellations.

The \textit{k}-Triangle-Faithfulness assumption is a weakening of the
Strong Causal Faithfulness assumption in two ways. First,
Triangle-Faithfulness is significantly weaker than Faithfulness.
Second, it does not entail a lower limit on the size of nonzero partial
correlations; it only puts a limit on the size of a nonzero partial
correlation in relation to the size of the structural coefficient of an
edge that occurs in a triangle.

The Strong Causal Faithfulness assumption entails that there are no
very small structural coefficients (which, if present, entail the
existence of some partial correlation that is very small). In contrast,
the \textit{k}-Triangle-Faithfulness assumption does not entail that there
are no nonzero but very small structural coefficients. However, there
is a price to be paid for weakening the Strong Causal Faithfulness
assumption; the estimator we propose is both computationally more
intensive than the PC algorithm used in \citet{KalBuh07}
and also requires testing partial correlations conditional on larger
sets of variables, which means some of the tests performed have lower
power than the tests performed in the PC algorithm.

Our results also depend on the following assumptions. First, we assume
a fixed upper bound to the size of the set of variables that does not
change as sample size increases. We have no reason to think that there
are not analogous results that would hold even if, as in \citet{KalBuh07}, the number of variables and the degree of the graph
increased with the sample size; however, we have not proved any such
results yet. We also make the assumption of nonvanishing variance (NVV)
and the assumption of upper bound for partial correlations (UBC):

\textit{Assumption NVV\textup{(}J\textup{)}}.
\begin{eqnarray}
\mathop{\inf}_{{{X}_{i}}\in\mathbf{V}} {{\operatorname {var}}_{M}}
\bigl({{X}_{i}}|\mathbf{V}\setminus\{{{X}_{i}}\}\bigr) \ge
J \nonumber\\
\eqntext{\mbox{for some (small) }J>0.}
\end{eqnarray}

\textit{Assumption UBC\textup{(}C\textup{)}}.
\begin{eqnarray}
\mathop{\sup}_{{{X}_{i}},{{X}_{j}}\in\mathbf{V},\mathbf{W}\subseteq
\mathbf
{V}\setminus\{{{X}_{i}},{{X}_{j}}\}} \bigl\llvert {{\rho }_{M}}({{X}_{i}},{{X}_{j}}|
\mathbf{W}) \bigr\rrvert \le C\nonumber\\
\eqntext{\mbox{for some }C< 1.}
\end{eqnarray}

The assumption NVV is a slight strengthening of the positivity
requirement, which, as we noted in the previous section, is needed to
guarantee the Causal Minimality assumption. Uniform consistency
requires that the distributions be bounded away from nonpositivity.

The assumption UBC [cf. the second part of assumption (A4)] is used to
guarantee that sample partial correlations are uniformly consistent
estimators of population partial correlations (\cite{KalBuh07}).

We now proceed to establish two positive results about uniform
consistency. In Section~\ref{sec4.1} we show that the \textit{Conservative SGS}
(\textit{CSGS}) algorithm, using uniformly consistent tests of partial
correlations, is uniformly consistent in inferring certain features of
the causal structure. In Section~\ref{sec4.2} we show that the \textit{Very
Conservative SGS} (\textit{VCSGS}) algorithm, when combined with a
uniformly consistent procedure for estimating structural coefficients,
provides a uniformly consistent estimator of structural coefficients
(that returns ``Unknown'' in some, but not all cases).

\subsection{Uniform Consistency in the Inference of Structure}\label{sec4.1}
Recall that the \textit{CSGS} algorithm, given a perfect oracle of
conditional independence, is correct under the Causal Markov,
Minimality and Triangle-Faithfulness assumptions, in the sense that the
adjacencies, orientations and marked noncolliders in the output are all
correct. In Gaussian models, we can implement the oracle with tests of
zero partial correlations. A test $\varphi$ of $H_0\dvtx \rho= 0$
versus $H_1\dvtx \rho\neq0$ is a family of functions: $\varphi
_1,\ldots,\varphi_n, \ldots,$ one for each sample size, that takes an
i.i.d. sample $\mathit{V}_n$ from the joint distribution over \textbf{V} and
returns 0 (acceptance of $H_0$) or 1 (rejection of $H_0)$. Such a
test is \textit{uniformly consistent} with respect to a set of
distributions $\Omega$ if and only if
\begin{longlist}[1.]
\item[1.]${\mathop{\lim}}_{n\to\infty} {\mathop{\sup}}_{P\in\Omega
\wedge\rho(P)=0} P_{{}}^{n}({{\varphi
}_{n}}({{V}_{n}})=1)=0$, and
\item[2.]${\mbox{for every }}\delta> 0$,
\[
\mathop{\lim} _{n \to
\infty} \mathop{\sup} _{P \in\Omega\wedge|\rho(P)| \geq
\delta} {P^n}\bigl({\varphi_n}({V_n}) = 0\bigr) = 0.
\]
\end{longlist}

For simplicity, we assume the variables in \textbf{V} are standardized.
Under the assumption UBC, there are uniformly consistent tests of
partial correlations based on sample partial correlations, such as
Fisher's \textit{z} test (\cite{Robetal03}; \cite{KalBuh07}). We consider a sample version of the \textit{CSGS} algorithm in
which the oracle is replaced by uniformly consistent tests of zero
partial correlations in the adjacency step S2. In the orientation
phase, the step S3* is refined as follows, based on a user chosen
parameter \textit{L}.

S3* (sample version). Let \textit{K} be the undirected graph resulting
from the adjacency phase. For each unshielded triple $ \langle X, Y,
Z\rangle$,
\begin{longlist}[1.]
\item[1.] If there is a set \textbf{W} not containing \textit{Y} such that the
test of $\rho(X,Z|\mathbf{W}) = 0$ returns 0 (i.e., accepts the
hypothesis), and for every set \textbf{U} that contains \textit{Y}, the test
of $|\rho(X,Z|\mathbf{U})| = 0$ returns 1 (i.e., rejects the
hypothesis), and the test of $|\rho(X,Z|\mathbf{U}) - \rho(X,Z|\mathbf{W})|
\geq L$ returns 0 (i.e., accepts the hypothesis), then orient the
triple as a collider: $X \rightarrow Y \leftarrow Z$.
\item[2.] If there is a set \textbf{W} containing \textit{Y} such that the test
of $\rho(X,Z|\mathbf{W}) = 0$ returns 0 (i.e., accepts the hypothesis),
and for every set \textbf{U} that does not contain \textit{Y}, the test of $
|\rho(X,Z|\mathbf{U})| = 0$ returns 1 (i.e., rejects the hypothesis), and
the test of $|\rho(X,Z|\mathbf{U}) - \rho(X,Z|\mathbf{W})| \geq L$ returns
0 (i.e., accepts the hypothesis), then mark the triple as a noncollider.
\item[3.] Otherwise, mark the triple as ambiguous.
\end{longlist}

Larger values of \textit{L} return ``Unknown'' more often than smaller
values of \textit{L}, but reduce the probability of an error in
orientation at a given sample size.

Step S4 remains the same as in the population version.

Given any causal model $M = \langle P, G\rangle$ over \textbf{V}, let $
C(L, n, M)$ denote the (random) output of the \textit{CSGS} algorithm
with parameter \textit{L}, given an i.i.d. sample of size \textit{n} from the
distribution $P_M$. Say that $C(L, n, M)$ errs if it contains (i)
an adjacency not in $G_M$, or (ii) a marked noncollider not in $G_M$, or (iii)
an orientation not in $G_M$.\footnote{Note that at this stage we are taking non-adjacencies as uninformative, and not
counting any missing edge as an error. So an algorithm that always returns a structure
with no edges is treated as totally uninformative and hence trivially consistent, in the
sense of triviality defined in Robins et al. (\citeyear{Robetal03}). The CSGS algorithm is obviously
nontrivial in that it does not always return a completely uninformative answer.}

Let $\psi^{k,J,C}$ be the set of causal models over \textbf{V} that
respect the \textit{k}-Triangle-Faithfulness assumption and the
assumptions of NVV(\textit{J}) and UBC(\textit{C}). We shall prove that given
the causal sufficiency of the measured variables \textbf{V} and the causal
Markov assumption,
\[
\mathop{\lim}_{n\to\infty} \mathop{\sup}_{M\in{{\psi
}^{k,J,C}}}
P_{M}^{n}\bigl(C(L,n,M)\ \mathrm{errs}\bigr)=0.
\]
In other words, given the causal sufficiency of \textbf{V}, the Causal
Markov, \textit{k}-Triangle-Faithfulness, NVV(\textit{J}) and UBC(\textit{C})
assumptions, the \textit{CSGS} algorithm is uniformly consistent in that
the probability of it making a mistake uniformly converges to zero in
the large sample limit.

First of all, we prove a useful lemma:
%
\begin{lemma}\label{le2}
Let $M \in\psi^{k,J,C}$. For any $X_i$ and $X_j$ such that $
X_j$ is not an ancestor of $X_i$, if $e_M(X_i \rightarrow X_j) =
b_{j,i}$, then
\[
\frac{\llvert  {{b}_{j,i}} \rrvert }{\sqrt{J}}\ge\bigl\llvert {{\rho }_{M}}\bigl(i,j|
\mathbf{X}[1,\ldots,j-1]\setminus\{{{X}_{i}}\}\bigr) \bigr\rrvert \ge
\llvert {{b}_{j,i}} \rrvert \sqrt{J},
\]
where $\mathbf{X}[1, \ldots, j]$ is an ancestral set that contains $
X_i$ but does not contain any descendant of $X_j$.
\end{lemma}
\begin{pf}
Let $\Sigma$ be the correlation matrix for the set of variables $\{
X_1,\ldots,X_j\}$, and $\mathbf{R} = \Sigma^{-1}$. Let \textbf{B} be the
(lower-triangular) matrix of structural coefficients in \textit{M}
restricted to $\{X_1,\ldots,X_j\}$, and var(\textbf{E}) be the
(diagonal) covariance matrix for the error terms $\{\varepsilon
_1,\ldots,\varepsilon_j\}$. Then
\[
\mathbf{R} = (\mathbf{I} {} - \mathbf{B})^{T}\operatorname{var}(
\mathbf{E})^{-1}(\mathbf{I} {} - \mathbf{B}).
\]
Note that
\begin{eqnarray*}
(\mathbf{I}-\mathbf{B})&=&\lleft[ %
\matrix{ 1 & 0 & \cdots& 0
\vspace*{2pt}
\cr
-{{b}_{2,1}} & 1 & \cdots& 0 \vspace*{2pt}
\cr
\vdots&
\cdots& \cdots& 0 \vspace*{2pt}
\cr
-{{b}_{j,1}} & \cdots&
-{{b}_{j,j-1}} & 1 } \rright]\operatorname{var}(\mathbf{E})^{-1}\\
&=&
\lleft[ %
\matrix{ 1/{{\varepsilon}_{1}} & 0 & \cdots& 0
\vspace*{2pt}
\cr
0 & 1/{{\varepsilon}_{2}} & \cdots& 0 \vspace*{2pt}
\cr
\vdots& \cdots& \cdots& 0 \vspace*{2pt}
\cr
0 & \cdots& 0 & 1/{{
\varepsilon}_{j}} } \rright],
\end{eqnarray*}
where the \textit{b}'s are the corresponding structural coefficients in
\textit{M}, and the $\varepsilon$'s are the variances of the
corresponding error terms. Thus, $\mathbf{R}[j, j] = 1/\varepsilon_j$,
and $\mathbf{R}[i, j] = -b_{j,i}/\varepsilon_{j}$. So we have \citepp{Whi90}
\begin{eqnarray*}
&&{{\rho}_{M}}\bigl({{X}_{i}},{{X}_{j}}|
\mathbf{X}[1,\ldots,j-1]\setminus\{ {{X}_{i}}\}\bigr)\\
&&\quad=-\frac{\mathbf{R}[i,j]}{{{ ( \mathbf{R}[i,i]\cdot
\mathbf{R}[j,j]  )}^{1/2}}}=
\frac{{{b}_{j,i}}}{\mathbf
{R}{{[i,i]}^{1/2}}\varepsilon_{j}^{1/2}}.
\end{eqnarray*}
Since $\mathbf{R}[i,i]^{- 1}$ is the variance of $X_i$ conditional on
all of the other variables in $\{X_1,\ldots,X_j\}$, which is a subset
of $\mathbf{V} \setminus\{X_i\}$, $\mathbf{R}[i,i]^{-1} \geq
\operatorname{var}_M(X_i|\mathbf{V} \setminus\{X_i\}) \geq J$. Since
the variables are standardized and the residual of $X_i$ regressed on
the other variables is uncorrelated with $X_i$, $\mathbf{R}[i,i]^{-1}
\leq1$. Similarly, $1 \geq\varepsilon_j \geq J$. Thus,
\[
\frac{\llvert  {{b}_{j,i}} \rrvert }{\sqrt{J}}\ge\bigl\llvert {{\rho }_{M}}\bigl(i,j|
\mathbf{X}[1,\ldots,j-1]\setminus\{{{X}_{i}}\}\bigr) \bigr\rrvert \ge
\llvert {{b}_{j,i}} \rrvert \sqrt{J}.
\]\hspace*{4pt}\upqed
\end{pf}

We now categorize the mistakes $C(L, n, M)$ can make into three kinds.
$C(L, n, M)$ \textit{errs in kind I} if $C(L, n, M)$ has an adjacency that
is not present in $G_M$; $C(L, n, M)$ \textit{errs in kind II} if every
adjacency in $C(L, n, M)$ is in $G_M$ but $C(L, n, M)$ contains a
marked noncollider that is not in $G_M$; $C(L, n, M)$ errs in \textit{kind III} if every adjacency in $C(L, n, M)$ is in $G_M$, every
marked noncollider in $C(L, n, M)$ is in $G_M$, but $C(L, n, M)$
contains an orientation that is not in $G_M$. Obviously if $C(L, n,
M)$ errs, it errs in at least one of the three kinds.

The following three lemmas show that for each kind, the probability of
$C(L, n, M)$ erring in that kind uniformly converges to zero.

\begin{lemma}\label{le3}
Given causal sufficiency of the measured variables $\mathbf{V}$, the Causal
Markov, \textit{k}-Triangle-Faithfulness, NVV(\textit{}J) and UBC(\textit{C})
assumptions,
\begin{eqnarray*}
&&\mathop{\lim}_{n\to\infty} \mathop{\sup}_{M\in{{\psi
}^{k,J,C}}}
P_{M}^{n}\bigl(C(L,n,M)\mbox{ errs in kind I}\bigr)\\
&&\quad=0.
\end{eqnarray*}
\end{lemma}
\begin{pf}
$C(L, n, M)$ has an adjacency not in $G_M$ only if some test of
zero partial correlation falsely rejects its null hypothesis. Since
uniformly consistent tests are used in \textit{CSGS}, for every $
\varepsilon> 0$, for every test of zero partial correlation $t_i$,
there is a sample size $N_i$ such that for all $n > N_i$ the
supremum (over $\psi^{k,J,C})$ of the probability of the test falsely
rejecting its null hypothesis is less than $\varepsilon$. Given \textbf{V}, there are only finitely many possible tests of zero partial
correlations. Thus, for every $\varepsilon> 0$, there is a sample
size \textit{N} such that for all $n > N$, the supremum (over $\psi
^{k,J,C})$ of the probability of any of the tests falsely rejecting
its null hypothesis is less than $\varepsilon$. The lemma then follows.
\end{pf}

\begin{lemma}\label{le4}
Given causal sufficiency of the measured variables $\mathbf{V}$, the Causal
Markov, \textit{k}-Triangle-Faithfulness, and NVV(\textit{J}) and UBC(\textit{C}) assumptions,
\[
\mathop{\lim} _{n \to\infty} \mathop{\sup} _{M \in
{\psi^{k,J,C}}} P_M^n
\bigl(C(L,n,M) \mbox{ errs in kind II}\bigr) = 0.
\]
\end{lemma}
\begin{pf}
For any $M \in\psi^{k,J,C}$, if $C(L, n, M)$ errs in kind II, then
$C(L, n, M)$ contains a marked noncollider, say, $\langle X, Y,
Z\rangle$ which is not in $G_M$, but every adjacency in $C(L, n,
M)$ is also in $G_M$, including the adjacency between \textit{X} and
\textit{Y}, and that between \textit{Y} and \textit{Z}. It follows that $
\langle X, Y, Z\rangle$ is a collider in $G_M$. Since \textit{CSGS}
marks a triple as a noncollider only if the triple is unshielded, \textit{X} and \textit{Z} are not adjacent in $C(L, n, M)$. Hence, errors of
kind II can be further categorized into two cases: (II.1) $C(L, n,
M)$ contains an unshielded noncollider that is an \textit{unshielded}
collider in $G_M$, and (II.2) $C(L, n, M)$ contains an unshielded
noncollider that is a \textit{shielded} collider in $G_M$. We show that
the probability of either case uniformly converges to zero.

For case (II.1) there is an unshielded collider $\langle X, Y, 
Z\rangle$ in $G_M$, so \textit{X} and \textit{Z} are independent
conditional on some set of variables \textbf{W} that does not contain \textit{Y}, by the Causal Markov assumption. Then the \textit{CSGS} algorithm
(falsely) marks $\langle X, Y, Z\rangle$ as a noncollider only if
the test of $\rho_M(X, Z|\mathbf{W}) = 0$ (falsely) rejects its null
hypothesis. Therefore, the \textit{CSGS} algorithm gives rise to case
(II.1) only if some test of zero partial correlation falsely rejects
its null hypothesis. Then, by essentially the same argument as the one
used in proving Lemma~\ref{le3}, the probability of case (II.1) uniformly
converges to zero as sample size increases.

For case (II.2), suppose for the sake of contradiction that the
probability of \textit{CSGS} making such a mistake does not uniformly
converge to zero. Then there exists $\varepsilon> 0$, such that for
every sample size \textit{n}, there is a model $M(n)$ such that the
probability of $C(L, n, M(n))$ contains an unshielded noncollider
that is a shielded collider in $M(n)$ is greater than $\varepsilon$.

Now, $C(L, n, M(n))$ contains an unshielded noncollider that is a
shielded collider in $G_{M(n)}$, say $\langle X^{M(n)}, Y^{M(n)},
Z^{M(n)}\rangle$, only if there is a set $\mathbf{W}^{M(n)}$ that
contains \textit{Y} such that the test of $\rho(X^{M(n)},\break Z^{M(n)}) |
\mathbf{W}^{M(n)}) = 0$ returns 0 (i.e., accepts the hypothesis).

Without loss of generality, suppose $Z^{M(n)}$ is not an ancestor of
$X^{M(n)}$. Let $\mathbf{U}^{M(n)} = \mathbf{A}^{M(n)}\setminus\{
X^{M(n)},  Z^{M(n)}\}$, where $\mathbf{A}^{M(n)}$ is an ancestral set
that contains $X^{M(n)}$ and $Z^{M(n)}$ but no descendent of $
Z^{M(n)}$. Since $Y^{M(n)}$ is a child of $Z^{M(n)}$ in $
G_{M(n)}$, $\mathbf{U}^{M(n)}$ does not contain $Y^{M(n)}$. Then, $
\langle X^{M(n)}, Y^{M(n)}, Z^{M(n)}\rangle$ is marked as a
noncollider in $C(L, n, M(n))$ only if the test of $|\rho(X^{M(n)},
Z^{M(n)} \break| \mathbf{U}^{M(n)}) - \rho(X^{M(n)}, Z^{M(n)} |  \mathbf{W}^{M(n)})|
\geq L$ returns 0 (i.e., accepts the hypothesis).

The test of $|\rho(X^{M(n)}, Z^{M(n)} | \mathbf{U}^{M(n)}) - \rho
(X^{M(n)},\break  Z^{M(n)} | \mathbf{W}^{M(n)})| \geq L$ will be denoted by $
\varphi_{n(L)}$ and $\varphi_{n(0)}$ denotes the test of $\rho
(X^{M(n)}, Z^{M(n)} | \mathbf{W}^{M(n}) = 0$. By our supposition,
$P_{M(n)}^{n}({{\varphi}_{n(0)}}=0$ and ${{\varphi
}_{n(L)}}=0)>\varepsilon$. It follows that for all \textit{n},
%
\begin{longlist}[(1)]
\item[(1)] $P_{M(n)}^n({\varphi
_{n(0)}} = 0) > \varepsilon$,
\item[(2)] $P_{M(n)}^n({\varphi_{n(L)}} = 0) > \varepsilon$.
\end{longlist}
%

(1) implies that there exists $\delta_n$ such that $|\rho
(X^{M(n)},\allowbreak  Z^{M(n)}|\mathbf{W}^{M(n})| < \delta_n$, and $\delta_n
\rightarrow0 $ as $n \rightarrow\infty$ since the tests are
uniformly consistent. $|e_M(X^{M(n)}\mbox{ ---}\linebreak[4]  Z^{M(n)})| \leq|\rho
(X^{M(n)}, Z^{M(n)} | \mathbf{W}^{M(n)})| / k < \delta_n/k$ by \textit{k}-Triangle-Faithfulness.
By Lemma \ref{le2}, $|\rho(X^{M(n)},\break Z^{M(n)} |  \mathbf{U}^{M(n)})| \leq J^{-1/2} |e_M(X^{M(n)} \mbox{ --- }Z^{M(n)})| <\break  \delta_n
J^{-1/2}/k$.

Thus, $|\rho(X^{M(n)}, Z^{M(n)} | \mathbf{U}^{M(n)}) - \rho
(X^{M(n)},  Z^{M(n)} |\linebreak[4]  \mathbf{W}^{M(n)})| < \delta_n(1+ J^{-1/2}/k)
\rightarrow0$ as $n \rightarrow\infty$. Therefore, it is not true
that (2) holds for all \textit{n}, which is a contradiction. So the
initial supposition is false. The probability of case (II.2) uniformly
converges to zero as sample size increases.
\end{pf}

\begin{lemma}\label{le5}
Given causal sufficiency of the measured variables $\mathbf{V}$, the Causal
Markov, \textit{k}-Triangle-Faithfulness, NVV(\textit{J}) and UBC(\textit{C})
assumptions,
\[
\mathop{\lim} _{n \to\infty} \mathop{\sup} _{M \in
{\psi^{k,J,C}}} P_M^n
\bigl(C(L,n,M)\mbox{ errs in kind III}\bigr) = 0.
\]
\end{lemma}
\begin{pf} Given that all the adjacencies and marked noncolliders in
$C(L, n, M)$ are correct, there is a mistaken orientation if and only
if there is an unshielded collider in $C(L, n, M)$ which is not a
collider in $G_M$, for the other orientation rules in step S4 would
not lead to any mistaken orientation if all the unshielded colliders
were correct. Thus, $C(L, n, M)$ errs in kind III only if there is a
noncollider $\langle X, Y, Z\rangle$ in $G_M$ that is marked as an
unshielded collider in $C(L, n, M)$.

There are then two cases to consider: (III.1) $C(L, n,  M)$ contains
an unshielded collider that is an \textit{unshielded} noncollider in $
G_M$, and (III.2) $C(L, n, M)$ contains an unshielded collider that
is a \textit{shielded} noncollider in $G_M$. The argument for case
(III.1) is extremely similar to that for (II.1) in the proof of Lemma~\ref{le4}, and the argument for case (III.2) is extremely similar to that for
(II.2) in the proof of Lemma~\ref{le4}.
\end{pf}

\begin{theorem}\label{th1} Given causal sufficiency of the measured variables $\mathbf{V}$, the Causal Markov, \textit{k}-Tri\-angle-Faithfulness, NVV(\textit{J}) and
UBC(\textit{}C) assumptions, the \textit{CSGS} algorithm is uniformly
consistent in the sense that
\[
\mathop{\lim} _{n \to\infty} \mathop{\sup} _{M \in
{\psi^{k,J,C}}} P_M^n
\bigl(C(L,n,M)\mbox{ errs}\bigr) = 0.
\]
\end{theorem}
\begin{pf}
It follows from Lemmas \ref{le3}--\ref{le5} [and the fact that $C(L, n, M)$ errs if
and only if it errs in one of the three kinds].
\end{pf}

\subsection{Uniform Consistency in the Inference of Structural Coefficients}\label{sec4.2}
We now combine the structure search with estimation of structural
coefficients, when possible.

\textit{Edge Estimation algorithm}.
\begin{enumerate}[E1.]
\item[E1.] Run the \textit{CSGS} algorithm on an i.i.d. sample of size
\textit{n} from $P_M$.
\item[E2.] Let the output from E1 be $C(L, n, M)$. Apply step V5 in
the \textit{VCSGS} algorithm (from Section~\ref{sec3}), using tests of zero partial
correlations.
\item[E3.] If the nonadjacencies in $C(L, n, M)$ are not confirmed in
E2, return ``Unknown'' for every pair of variables.
\item[E4.] If the nonadjacencies in $C(L, n, M)$ are confirmed in E2, then
\begin{enumerate}[(iii)]
\item[(i)] For every nonadjacent pair $\langle X, Y\rangle$, let the
estimate $\hat e(X\mbox{ --- }Y)$ be 0.
\item[(ii)]For each vertex \textit{Z} such that all of the edges
containing \textit{Z} are oriented in $C(L, n, M)$, if \textit{Y} is a
parent of \textit{Z} in $C(L, n, M)$, let the estimate $\hat e(Y\mbox{ --- }
Z)$ be the sample regression coefficient of \textit{Y} in the regression
of \textit{Z} on its parents in $C(L, n, M)$.
\item[(iii)] For any of the remaining edges, return ``Unknown.''
\end{enumerate}
\end{enumerate}

The basic idea is that we first run the Very Conservative SGS (\textit{VCSGS}) algorithm, which, recall, is the \textit{CSGS} algorithm (E1) plus
a step of testing whether the output satisfies the Markov condition
(E2). If the test does not pass, we do not estimate any edge; if the
test passes, we estimate those edges that are into a vertex that is not
part of any unoriented edge.

Let $M_1$ be an output of the Edge Estimation algorithm, and $M_2$
be a causal model. We define the \textit{structural coefficient distance},
$d[M_1,M_2]$, between $M_1$ and $M_2$ to be
\begin{eqnarray*}
&&d[{M_1},{M_2}] \\
&&\quad= \mathop{\max} _{i,j} \bigl
\llvert {{{\hat e}_{{M_1}}}({X_i} \to{X_j}) -
{e_{{M_2}}}({X_i} \to{X_j})}\bigr\rrvert ,
\end{eqnarray*}
where by convention $\llvert  {{\hat e}_{{M_1}}}({X_i} \to{X_j}) -
{e_{{M_2}}}({X_i} \to {X_j}) \rrvert  = 0$ if ${\hat e_{{M_1}}}({X_i}
\to{X_j}) = {}$``Unknown.''

Intuitively, the structural coefficient distance between the output and
the true causal model measures the (largest) estimation error the Edge
Estimation algorithm makes. Our goal is to show that under the
specified assumptions, the Edge Estimation algorithm is uniformly
consistent, in the sense that for every $\delta> 0$, the probability
of the structural coefficient distance between the output and the true
model being greater than $\delta$ uniformly converges to zero.

Obviously, by placing no penalty on the uninformative answer of
``Unknown,'' there is a trivial algorithm that is uniformly consistent,
namely, the algorithm that always returns ``Unknown'' for every
structural coefficient. For this reason, \citet{Robetal03} also
requires any admissible algorithm to be nontrivial in the sense that it
returns an informative answer (in the large sample limit) for some
possible joint distributions. The Edge Estimation algorithm is clearly
nontrivial in this sense. There is no guarantee that it will always
output an informative answer for some structural coefficient, and
rightly so, because there are cases---for example, when the true
causal graph is a complete one and there is no prior information about
the causal order---in which every structural coefficient is truly
underdetermined or unidentifiable. An interesting question, however, is
whether a given algorithm is maximally informative or complete in the
sense that it returns (in the large sample limit) ``Unknown'' only on
those structural coefficients that are truly underdetermined. The
condition in question is of course much stronger than Robins et al.'s
condition of nontriviality. We suspect that the Edge Estimation
algorithm is not maximally informative in this sense. (We thank an
anonymous referee for raising this issue.)

\begin{theorem}\label{th2} Given causal sufficiency of the measured variables $\mathbf{V}$, the Causal Markov, \textit{k}-Triangle-Faithfulness, NVV(\textit{J}) and
UBC(\textit{C}) assumptions, the Edge Estimation algorithm is uniformly
consistent in the sense that for every $\delta > 0$
\[
\mathop{\lim} _{n \to\infty} \mathop{\sup} _{M \in
{\psi^{k,J,C}}} P_M^n
\bigl(d\bigl[\hat O(M),M\bigr] > \delta\bigr) = 0,
\]
where $\hat O(M)$ is the output of the algorithm given an i.i.d. sample
from $P_M$.
\end{theorem}

\begin{pf} Let \textbf{O} be the set of possible graphical outputs of
the \textit{CSGS} algorithm. Given \textbf{V}, there are only finitely many
graphs in \textbf{O}. So it suffices to show that for each $O \in\mathbf{O}$,
\begin{eqnarray*}
&&\mathop{\lim} _{n \to\infty} \mathop{\sup} _{M \in
{\psi^{k,J,C}}} P_M^n
\bigl(d\bigl[\hat O(M),M\bigr] > \delta|\\
&& \hspace*{88pt}C(L,n,M) = O\bigr)\\
&&\quad\hspace*{51pt}{}\cdot P_M^n
\bigl(C(L,n,M) = O\bigr) = 0.
\end{eqnarray*}
Given \textit{O}, $\psi^{k,J,C}$ can be partitioned into the following
three sets:
\begin{eqnarray*}
\Psi_{1} &= &\{\textit{M}|\mbox{All } \mbox{adjacencies}, \mbox{ nonadjacencies }
\mbox{and}\\
&&\hspace*{44pt}\mbox{orientations } \mbox{in } \textit{O} \mbox{ are } \mbox{true } \mbox{of } \textit{M}\};
\\
\Psi_{2}& =& \{\textit{M}|\mbox{\textit{O} contains an adjacency or
an}\\
&&\hspace*{40pt}\mbox{orientation not true of \textit{M}}\};\\
\Psi_{3} &=& \{\textit{M}|\mbox{All adjacencies and orientations in
\textit{O} are} \\
&&\hphantom{\{\textit{M}|}\mbox{true of \textit{M}, but some nonadjacencies are}\\
&&\hspace*{137pt}\mbox{not true of
\textit{M}}\}.
\end{eqnarray*}

It suffices to show that for each $\Psi_{i}$,
\begin{eqnarray*}
&&\mathop{\lim} _{n \to\infty} \mathop{\sup} _{M \in
{\psi_i}} P_M^n
\bigl(d\bigl[\hat O(M),M\bigr] > \delta|C(L,n,M) = O\bigr)\\
&&\hspace*{48pt}{}\cdot P_M^n
\bigl(C(L,n,M) = O\bigr) = 0.
\end{eqnarray*}
Consider $\Psi_{1}$ first. Given any $M \in\Psi_1$, the zero
estimates in $\hat O(M)$ are all correct (since all nonadjacencies are
true). For each edge $Y \rightarrow Z$ that is estimated, the true
structural coefficient $e_M(Y \rightarrow Z)$ is simply $
r_M(Y,Z,\mathbf{Parents}(O,Z))$, the population regression coefficient
for \textit{Y} when \textit{Z} is regressed on its parents in \textit{O},
because the set of \textit{Z}'s parents in \textit{O} is the same as the set
of \textit{Z}'s parents in $G_M$.

The sampling distribution of the estimate of an edge $X \rightarrow
Y$ in $O$ is given by
%
%
\begin{eqnarray*}
&&{\hat r_M}\bigl(Y,Z,{\mathbf{Parents}}(O,Z),n\bigr)
\\
&&\quad\sim\mathcal{N} \biggl( {r_M}\bigl(Y,Z,{\mathbf{Parents}}(O,Z)\bigr),
\\
&&\hspace*{41pt}{}\frac{{\sigma_e^2}}{{n\operatorname
{var} ( Y \rrvert {\mathbf{Parents}}(O,Z)\setminus\{ Y\} )}} \biggr),
\end{eqnarray*}
where $\sigma_e^2$ is the variance of the residual for \textit{Z} when
regressed upon $\mathbf{Parents}(O,Z)$ in $P_M$, and $\operatorname
{var} ( Y \rrvert  {\mathbf{Parents}}(O, Z)\setminus\{ Y\} )$
is the variance of \textit{Y} conditional on $\mathbf{Parents}(O,Z)\setminus Y$ in $P_M$ \citepp{Whi90}. The numerator
of the variance is bounded above by 1, since the variance of each
variable is 1, and the residual is independent of the set of variables
regressed on. The denominator is bounded away from zero by assumption
NVV(\textit{J}). Hence, sample regression coefficients are uniformly
consistent estimators of population regression coefficients under our
assumptions, and we have
%
%
\begin{eqnarray*}
&&\mathop{\lim} _{n \to\infty} \mathop{\sup} _{M \in
{\psi_1}} P_M^n
\bigl(d\bigl[\hat{O} (M),M\bigr] > \delta|C(L,n,M) = O\bigr)\\
&&\hspace*{46pt}{}\cdot P_M^n
\bigl(C(L,n,M) = O\bigr)
\\
&&\quad\leq\mathop{\lim} _{n \to\infty} \mathop{\sup} _{M
\in{\psi_1}}
P_M^n\bigl(d\bigl[\hat{O} (M),M\bigr] > \delta|\\
&&\hspace*{96pt} C(L,n,M)
= O\bigr) \\
&&\quad = 0.
\end{eqnarray*}
For $\Psi_2$, note that given any $M \in\Psi_2$, the \textit{CSGS}
algorithm errs if it outputs \textit{O}. Thus, by Theorem \ref{th1},
%
%
\begin{eqnarray*}
&&\mathop{\lim} _{n \to\infty} \mathop{\sup} _{M \in
{\psi_2}} P_M^n
\bigl(d\bigl[\hat{O} (M),M\bigr] > \delta|C(L,n,M) = O\bigr)\\
&&\hspace*{47pt}{}\cdot P_M^n
\bigl(C(L,n,M) = O\bigr)
\\
&&\quad\leq\mathop{\lim} _{n \to\infty} \mathop{\sup} _{M
\in{\psi_2}}
P_M^n\bigl(C(L,n,M) = O\bigr) = 0.
\end{eqnarray*}

Now consider $\Psi_3$. Let $O(M)$ be the population version of
$\hat O(M)$, that is, all the sample regression coefficients in $\hat O(M)$
are replaced by the corresponding population coefficients. Since sample
regression coefficients are uniformly consistent estimators of
population regression coefficients under our assumptions, and there are
only finitely many regression coefficients to consider, for every $
\varepsilon> 0$, there is a sample size $N_1$, such that for all $
n > N_1$, and all $M \in\Psi_3$,
\[
P_M^n \bigl( {d\bigl[\hat{O} (M),O(M)\bigr] >
\delta/2|C(L,n,M) = O} \bigr) < \varepsilon.
\]
For any $M \in\Psi_3$, there are some edges in $G_M$ missing in
\textit{O}. Let $\mathbf{E}(M)$ be the set of edges missing in \textit{O}. Let
$M'$ be the same as \textit{M} except that the structural coefficients
associated with the edges in $\mathbf{E}(M)$ are set to zero. Let $
O(M')$ be the same as $O(M)$ except that for each edge with an
identified coefficient, the coefficient in $O(M')$ is the relevant
regression coefficient derived from $P_M'$ [whereas that in $O(M)$
is derived from $P_M]$. By the setup of $M'$, the identified edge
coefficients in $O(M')$ are equal to the corresponding edge
coefficients in $M'$, which are the same as the corresponding edge
coefficients in $M$. Thus, the structural coefficient distance
between $O(M')$ and $M$ is simply
\[
d\bigl[O\bigl(M'\bigr),M\bigr] = \mathop{\max} _{ \langle i,j \rangle\in{\mathbf
{E}}(M)}
\bigl\llvert {{e_M}({X_i} \to{X_j})} \bigr
\rrvert .
\]
For any edge $Y\rightarrow Z$ in $O$ that has a different edge
coefficient in $O(M)$ than that in $O(M')$, the edge coefficients
are both derived from a regression of \textit{Z} on $\mathbf{Parents}(O,Z)$, but one is based on $P_M$ and the other is based on $P_{M'}$. The
regression coefficient $r(Y,Z, \mathbf{Parents}(O,Z))$ is equal to the
\textit{Y} component of the vector $\operatorname{cov}(Z, \mathbf{Parents}(O,Z))\*\operatorname{var}^{- 1}(\mathbf{Parents}(O, Z))$
\citepp{Whi90}, which, given the structure $G_M$, is a rational
function of the structural coefficients in \textit{M}. Since $M \in\psi
^{k,J,C}$, every submatrix of the covariance matrix for $P_M$ is
invertible, and so $r_M(Y,Z,  \mathbf{Parents}(O,Z))$ is defined. For $
M'$, $r_{M'}(Y,Z, \mathbf{Parents}(O,Z)) = r_{M'}(Y,Z, \mathbf{A})$, where
\textbf{A} is the smallest ancestral set that contains $\mathbf{Parents}(O,Z)$ in $G_M$.
$\operatorname{var}(\mathbf{A})^{ - 1} = (\mathbf{I} - \mathbf{B})^{T}
\operatorname{var}(\mathbf{E})^{- 1}(\mathbf{I} - \mathbf{B})$, where
\textbf{B} is the submatrix of structural coefficients in $M'$ for
variables in \textbf{A}, and $\operatorname{var}(\mathbf{E})$ is the
diagonal covariance matrix of error terms for variables in \textbf{A},
which is a submatrix of $\Sigma_M$. Since $M \in\psi^{k,J,C}$, the
variance of every error term is bounded from below by \textit{J}. Thus, $
\operatorname{var}(\mathbf{A})^{ - 1}$\linebreak[4]  is defined and so is $r_{M'}(Y,Z,
\mathbf{Parents}(O,Z))$.\linebreak[4]  Therefore, $r_M(Y,Z, \mathbf{Parents}(O,Z))$ and
$r_{M'}(Y,Z,\break \mathbf{Parents}(O,Z))$ are values of a rational function
of the structural coefficients.

A continuous function is uniformly continuous on a closed, bounded
interval anywhere that it is defined. A~rational function is continuous
at every point of its domain where its denominator is not zero, that 
is, where the function value is defined. By Lemma~\ref{le2} and assumption
UBC(\textit{C}), every structural coefficient $b_{j,i}$ in \textit{M} lies
in the closed bounded interval from $-C/J^{1/2}$ to $C/J^{1/2}$.
Obviously the coefficients in $M'$ still lie in this interval. Hence,
given $G_M$, the difference between $r_{M'}(Y,Z, \mathbf{Parents}(O,\break Z))$ and $r_M(Y,Z,   \mathbf{Parents}(O,Z))$ can be
arbitrarily small if the differences between the structural
coefficients in $M'$ and those in \textit{M} are sufficiently small.
Given the set of variables \textbf{V}, there are only finitely many
structures and finitely many relevant regressions to consider.
Therefore, there is a $\gamma\in(0, \delta/4)$ such that for every
$M \in\psi_3, d[O(M), O(M')] < \delta/4$ if
\[
\mathop{\max} _{ \langle i,j \rangle\in{\mathbf{E}}(M)} \bigl\llvert {{e_M}({X_i}
\to{X_j})} \bigr\rrvert < \gamma.
\]
Consider then the partition of $\Psi_3$ into
%
%
\[
{\Psi_{3.1}} = \Bigl\{ M \in{\Psi_3}\bigl|\mathop{\max}
_{ \langle
i,j \rangle\in{\mathbf{E}}(M)} \bigl\llvert {{e_M}({X_i}
\to{X_j})} \bigr\rrvert < \gamma\Bigr\}
\]
and
\[
{\Psi_{3.2}} = \Bigl\{ M \in{\Psi_3}\bigl|\mathop{\max}
_{ \langle
i,j \rangle\in{\mathbf{E}}(M)} \bigl\llvert {{e_M}({X_i}
\to{X_j})} \bigr\rrvert \geq\gamma\Bigr\} .
\]
It follows from the previous argument that for every $M \in\Psi
_{3.1}, d[O(M), M] \leq d[O(M), O(M')] + d[O(M'), M] < \delta/4+
\gamma< \delta/2$.
Then there is a sample size $N_1$, such that for all $n > N_1$ and
all $M \in\Psi_{3.1}$,
%
%
\begin{eqnarray*}
&&P_M^n \bigl( {d \bigl[ {\hat O(M),M} \bigr] >
\delta|C(L,n,M) = O} \bigr)
\\
&&\quad \leq P_M^n \bigl( {d \bigl[ {\hat O(M),O(M)} \bigr] >
\delta /2|C(L,n,M) = O} \bigr) \\
&&\quad< \varepsilon.
\end{eqnarray*}

For every $M \in\Psi_{3.2}$, there is at least one edge, say, $X
\rightarrow Y$ missing from \textit{O} such that $|e_M(X \rightarrow Y)|
\geq\gamma$. Then by Lemma \ref{le2}, there is a set \textbf{U} such that $|\rho
(X, Y | \mathbf{U})| \geq\gamma J^{1/2}$, but \textit{O} entails that $\rho
(X, Y |\mathbf{U}) = 0$. Thus, the test of the Markov condition in step
E2 is passed only if the test of $\rho(X, Y | \mathbf{U}) = 0$ returns 0
(i.e., accepts the null hypothesis). Note that if the test is not
passed, then every structural coefficient is ``Unknown,'' and so by
definition the structural coefficient distance is zero. Therefore, the
distance is greater than $\delta$ (and so nonzero) only if the test
of $\rho(X, Y | \mathbf{U}) = 0$ returns 0 while $|\rho(X, Y | \mathbf{U})| \geq\gamma J^{1/2}$. Since tests are uniformly consistent, it
follows that there is a sample size $N_2$, such that for all $n >
N_2$ and all $M \in\Psi_{3.2}$,
\[
P_M^n \bigl( {d \bigl[ {\hat O(M),M} \bigr] >
\delta|C(L,n,M) = O} \bigr) < \varepsilon.
\]
Let $N=\max(N_1, N_2)$. Then for all $n > N$,
%
%
\begin{eqnarray*}
&&\mathop{\sup} _{M \in{\psi_3}} P_M^n\bigl(d\bigl[\hat{O}
(M),M\bigr] > \delta|C(L,n,M) = O\bigr)\\
&&\hspace*{24pt}{}\cdot P_M^n
\bigl(C(L,n,M) = O\bigr)
\\
&&\quad\leq\mathop{\sup} _{M \in{\psi_3}} P_M^n\bigl(d\bigl[
\hat{O} (M),M\bigr] > \delta|C(L,n,M) = O\bigr) \\
&&\quad< \varepsilon.
\end{eqnarray*}
\upqed\end{pf}

\section{Conclusion}\label{sec5}
We have shown that there is a pointwise consistent estimator of causal
patterns and a uniformly consistent estimator of some of the structural
coefficients in causal patterns, even when the Causal Faithfulness
assumption and Strong Causal Faithfulness assumptions are substantially
weakened. The \textit{k}-Triangle Faithfulness assumption is a restriction
on many fewer partial correlations than the Causal Faithfulness
assumption and the Strong Causal Faithfulness assumptions, and does not
entail that there are no edges with very small but nonzero structural
coefficients.

There are a number of open problems associated with the Causal
Faithfulness assumption:

\begin{longlist}[1.]
\item[1.] Is it possible to speed up the \textit{Very Conservative SGS}
algorithm to make it applicable to data sets with large numbers of variables?
\item[2.] If unfaithfulness is detected, is it possible to reduce the
number of structural coefficients where the algorithm returns ``Unknown?''
\item[3.] In practice, on realistic sample sizes, how does the \textit{Very
Conservative SGS} algorithm perform? [\citet{RamZhaSpi06}, have already
shown that the \textit{Conservative PC} algorithm is more accurate and not
significantly slower than the \textit{PC} algorithm].
\item[4.] Is the \textit{k}-Triangle Faithfulness assumption unlikely to hold
for reasonable values of \textit{k} and large numbers of variables?
\item[5.] Is there an assumption weaker than the \textit{k}-Triangle
Faithfulness assumption for which there is a uniformly consistent
estimator for structural coefficients in a causal pattern?
\item[6.] Are there analogous results that apply when the number of
variables and the maximum degree of a vertex increases and the size of
\textit{k} decreases with increasing sample size [as in the \citet{KalBuh07}, results]?
\item[7.] Are there analogous results that apply when the assumption of
causal sufficiency is abandoned?
\item[8.] Are there analogous results that apply for other families of
distributions or for nonparametric tests of conditional independence?
\end{longlist}

\section*{Acknowledgments}
We thank two anonymous referees for helpful comments. Zhang's research
was supported in part by the Research grants Council of Hong Kong under
the General Research Fund LU341910.\vspace*{12pt}




\end{document}